\documentclass[]{ceurart}

\usepackage{listings}
\usepackage{graphicx}

\begin{document}

\copyrightyear{2026}
\copyrightclause{Copyright for this paper by its authors.
  Use permitted under Creative Commons License Attribution 4.0
  International (CC BY 4.0).}

\conference{LASI Spain 26: Learning Analytics Summer Institute Spain 2026, Zamora, May 07--08, 2026}

\title{AICoFe: Implementation and Deployment of an AI-Based Collaborative Feedback System for Higher Education}

\author[1]{Alvaro Becerra}[%
orcid=0009-0003-7793-2682,
email=alvaro.becerra@uam.es,
] \cormark[1]
\author[1]{Alejandra Palma}[%
orcid=0009-0005-6555-7250,
email=alejandra.palma@uam.es,
]
\author[1]{Ruth Cobos}[%
orcid=0000-0002-3411-3009,
email=ruth.cobos@uam.es,
]

\address[1]{GHIA Group, School of Engineering, Universidad Autónoma de Madrid, Spain}
\cortext[1]{Corresponding author.}

\begin{abstract}
Effective peer feedback is essential for developing critical reflection in higher education, yet its impact is often limited by the inconsistent quality of student-generated comments. This paper presents the implementation and deployment of AICoFe (AI-based Collaborative Feedback), a system designed to bridge this gap through a human-centered AI approach. We describe a modular architecture that orchestrates a multi-LLM pipeline—utilizing GPT-4.1-mini, Gemini 2.5 Flash, and Llama 3.1—to synthesize quantitative rubric data and qualitative observations into coherent, actionable feedback. Key to the system is a "teacher-in-the-loop" mediation workflow, where educators use specialized Learning Analytics dashboards to curate and refine AI-generated drafts before delivery. Furthermore, we detail the underlying data infrastructure, which employs a hybrid SQL and MongoDB strategy to ensure traceability and manage semi-structured feedback versions.
\end{abstract}


\maketitle

\section{Introduction}

Among the pedagogical strategies aimed at supporting the development of skills, peer feedback and peer assessment have proven to be particularly effective. By engaging students in the evaluation of their peers' work, peer feedback promotes active learning, critical reflection, and deeper understanding of quality criteria~\cite{liu2006peer,topping2009peer}. Moreover, acting as evaluators allows students not only to receive feedback but also to learn from the assessment process itself, strengthening their self-regulation and evaluative judgment skills~\cite{vanpopta2017exploring}.

Despite its educational potential, the implementation of peer feedback presents several well-documented challenges. Students often struggle to provide constructive, specific, and actionable comments, especially when they lack feedback literacy or confidence in their evaluative abilities~\cite{wei2024incorporating}. As a result, peer feedback may become superficial, inconsistent, or difficult for recipients to interpret and apply~\cite{viberg2024exploring}. These limitations can reduce student engagement and diminish the impact of peer assessment on learning outcomes.

Recent advances in Generative Artificial Intelligence (GenAI) open new opportunities to address these challenges and enhance peer feedback processes. Prior approaches to automatic feedback relied mainly on predefined rules, templates, or numerical indicators~\cite{bodily2018design,cavalcanti2021automatic}. More recently, large language models have demonstrated the ability to generate coherent, context-aware, and personalized feedback based on both quantitative data and qualitative inputs~\cite{giannakos2024promise}. In educational settings, GenAI has been successfully applied to improve the clarity, usefulness, and actionability of feedback, supporting students in understanding their performance and identifying concrete steps for improvement~\cite{kloos2024generative,zhang2024students}.

Building on this perspective, this paper presents AICoFe (\textit{Artificial Intelligence-based Collaborative Feedback} system), a system designed to enhance peer and self-evaluation processes in educational settings. From a human--computer interaction perspective, AICoFe is conceived as a human-centered, interactive system that supports human--AI collaboration through role-specific dashboards and teacher-in-the-loop feedback mediation. The system enables both teachers and students to conduct evaluations using customized rubrics that integrate quantitative scores and qualitative observations. Powered by generative AI, AICoFe generates personalized feedback for each student by combining the outputs of multiple large language models, providing actionable insights to improve their skills. The system's Learning Analytics (LA) dashboards offer dynamic features, including personalized summaries, comparative graphs, tools for self-assessment, and a feedback history interface that tracks LLM contributions and supports teacher draft management. Additionally, video recordings of student performances support reflective learning and a more thorough evaluation process.

Extending our previous work~\cite{becerra2025enhancing}, this paper focuses on the implementation details and deployment experience of the system. The remainder is structured as follows. Section~\ref{sec:related} briefly reviews related work. Section~\ref{sec:system} describes the system architecture and its main technical components in detail. Section~\ref{sec:conclusions} presents conclusions and future work.

\section{Related Work}
\label{sec:related}

\subsection{Automatic Feedback Systems in Education}

Before the emergence of GenAI, automatic feedback systems in education primarily
relied on rule-based mechanisms, comparing student responses to predefined answers
or visualising performance indicators through dashboards~\cite{bodily2018design,cavalcanti2021automatic}.
Providing automated feedback for oral presentations has been a notable application
of this line of work, with multimodal systems demonstrating measurable improvements
in presentation skills in real learning settings~\cite{ochoa2020controlled} and
recent open-source tools enabling broader adoption and contextual
adaptation~\cite{ochoa2024openopaf}. Some approaches incorporated machine learning
techniques such as embedding models to assess idea novelty in collaborative
settings~\cite{ulhaq2024novelty}, or used fixed templates to deliver automatic
feedback in online learning environments~\cite{topali2024instructor}.

\subsection{Learning Analytics Dashboards}

Learning Analytics dashboards (LADs) have emerged as a mechanism to support
reflection and self-regulation in educational settings, offering visualisations of
performance data that help students and teachers make sense of assessment
outcomes~\cite{verbert2013learning}. Research on LAD design has highlighted the
importance of aligning dashboard features with pedagogical goals and the specific
needs of different user roles, including students and instructors~\cite{bodily2018design}. Co-design approaches have proven
particularly effective in this regard, as illustrated by dashboards developed in
collaboration with instructors to deliver contextualised, human-centred feedback
in MOOC environments~\cite{topali2024instructor}. Related work has further expanded this perspective by exploring human-centered visual attention analytics based on eye-tracking and AI in online learning contexts \cite{navarro2024vaad,becerra2025integrating}, as well as multimodal dashboard systems for MOOCs \cite{becerra2023m2lads} that combine biometric, behavioral, and log data  to support a more holistic interpretation of learner activity~\cite{becerra2025enhancing_biosensors}.

\subsection{GenAI and LLMs for Feedback Generation}

The emergence of large language models has substantially expanded the possibilities
for automatic feedback generation in education. Unlike earlier rule-based
approaches, LLMs can produce coherent, context-aware, and personalised feedback
from both quantitative and qualitative inputs~\cite{giannakos2024promise}.
Empirical studies report positive student perceptions of GenAI-generated feedback,
particularly valuing its clarity, level of detail, and
usefulness~\cite{kloos2024generative,steiss2024comparing,wan2024exploring,zhang2024students}.

GenAI has also begun to support peer feedback specifically, helping to make
comments more constructive and actionable while reducing unhelpful or inappropriate
contributions~\cite{magyar2020balancing,sajadi2024harnessing,yang2025understanding}.
Peer assessment research has long highlighted both the potential and the limitations
of student-generated feedback: while peer feedback promotes active learning and evaluative judgment~\cite{liu2006peer,topping2009peer}, students frequently
struggle to produce constructive and specific comments, particularly when they lack feedback literacy~\cite{wei2024incorporating}, resulting in feedback that may be superficial, inconsistent, or difficult to act
upon~\cite{viberg2024exploring}. GenAI-assisted approaches offer a promising path to address these limitations at scale. Unlike prior systems that rely on a single model or fixed templates~\cite{topali2024instructor,ulhaq2024novelty}, AICoFe combines three independently fine-tuned LLMs under teacher mediation, enabling richer perspective diversity and preserving pedagogical authority over the final feedback.

\section{System Description}
\label{sec:system}

\begin{figure}[h]
  \centering
  \includegraphics[width=0.7\linewidth]{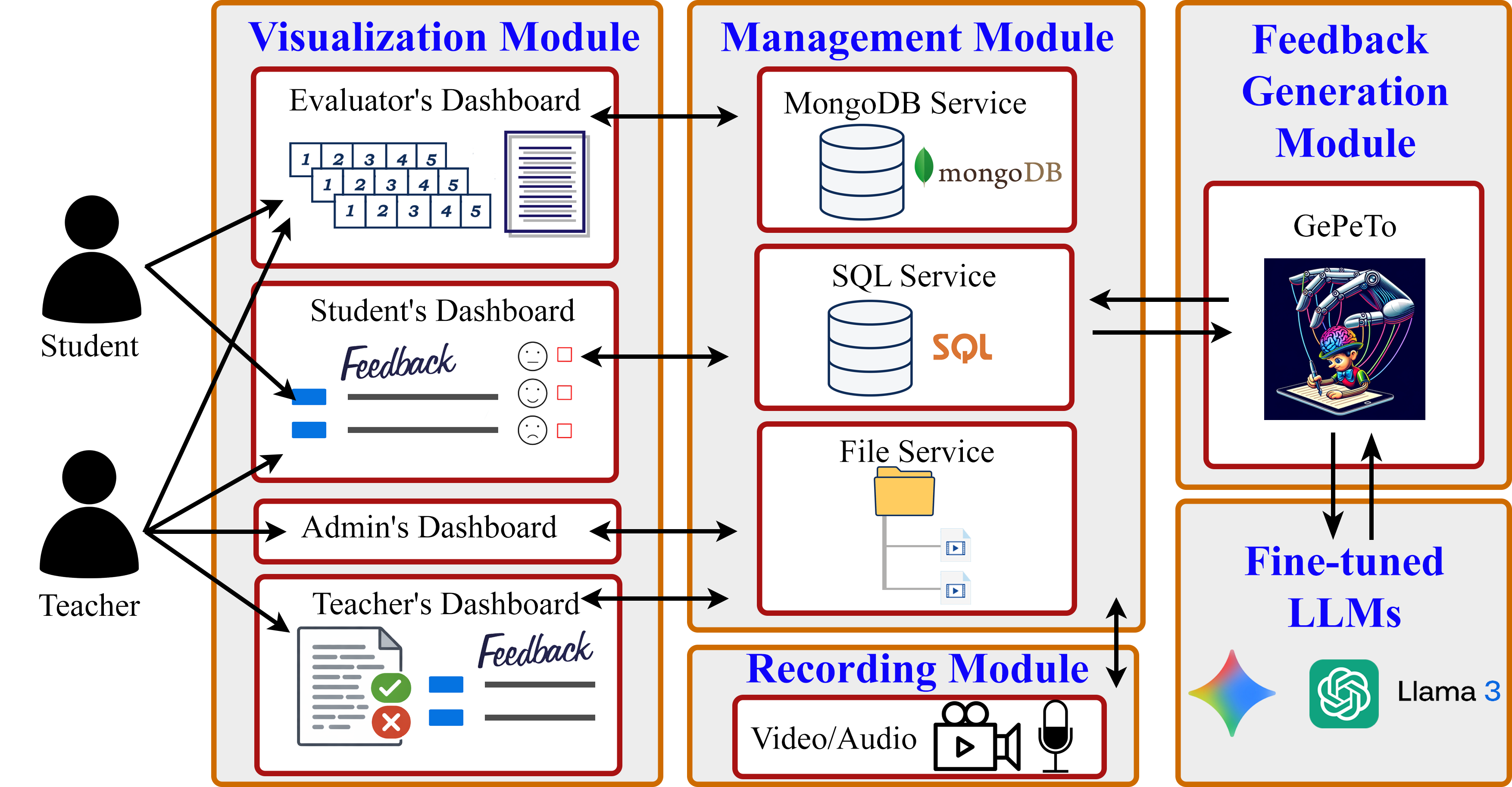}
  \caption{Architecture of the AICoFe system, showcasing its modular design with four main modules.}
  \label{fig:architecture}
\end{figure}

AICoFe is organised into four main modules (see Figure~\ref{fig:architecture}): (i) the \textbf{Visualization Module}, which provides role-specific dashboards; (ii) the \textbf{Management Module}, which handles data storage and connectivity; (iii) the \textbf{Feedback Generation Module}, which orchestrates multi-LLM feedback production; and (iv) the \textbf{Recording Module}, which captures student performances in video and audio format.

\subsection{Management Module}

The Management Module provides the data infrastructure underpinning the entire system. It integrates two complementary database technologies: a relational SQL database and a MongoDB document store.

The \textbf{SQL database} manages structured academic data, including courses, student groups, rubric definitions, and user accounts. It also stores the evaluators' quantitative scores and qualitative comments associated with each assessment. It supports the organisation of evaluation activities across multiple courses and cohorts, enabling the system to be deployed simultaneously in different educational contexts.

The \textbf{MongoDB database} is used to store semi-structured data that varies in shape across evaluation instances, such as the multiple feedback versions generated by each LLM, teacher drafts and curations, and the metadata associated with each feedback sentence (including its source LLM). It also stores the ratings of coherence and usefulness assigned to the generated feedback. This flexible document model is particularly suited to the multi-LLM feedback pipeline, where each evaluation instance may generate a variable number of feedback versions with different internal structures.

The \textbf{File Service} manages the storage and retrieval of audiovisual recordings associated with student presentations, linking them to the corresponding evaluation instances in the relational database. Prior to participation, students are provided with an informed consent form covering data usage and recording authorisation; participation in video recording is strictly opt-in. All audiovisual data are stored securely in compliance with European data protection regulations (GDPR).

\subsection{Visualization Module}

The Visualization Module provides a set of interactive Learning Analytics dashboards developed with the Dash framework. Each dashboard is tailored to the role of the user interacting with the system.

\subsubsection{Evaluator Dashboard}
\begin{figure}[t]
  \centering
  \includegraphics[width=0.8\linewidth, height=8cm, keepaspectratio=false]{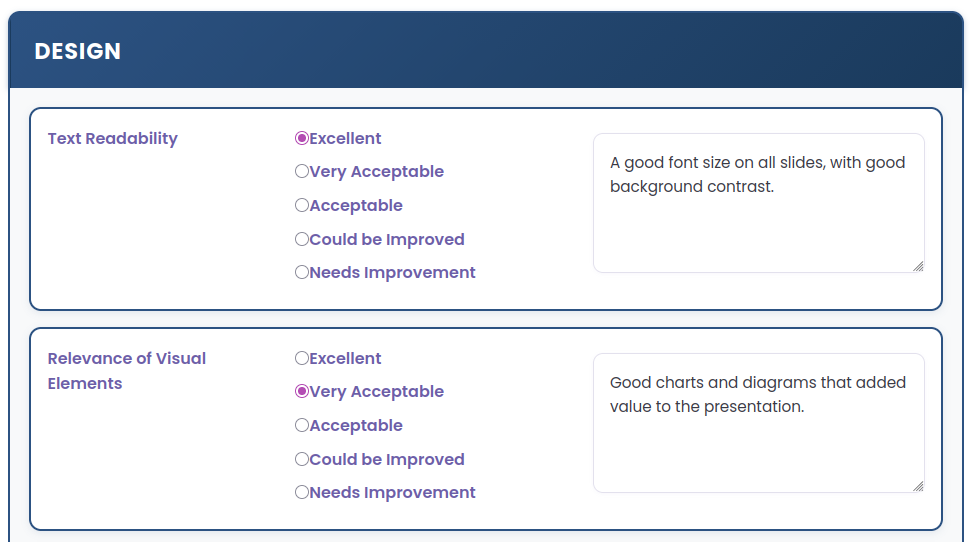}
  \caption{Screenshot of the AICoFe evaluator dashboard.}
  \label{fig:ev_dashboard}
\end{figure}

The Evaluator Dashboard (Figure~\ref{fig:ev_dashboard}) supports both teachers and students during the assessment process. After selecting a rubric and a target student, evaluators assign scores to each rubric item using a Likert scale and complement these with written qualitative comments. Detailed descriptions of the rubric levels remain accessible throughout the evaluation to support alignment in the interpretation of assessment criteria. The system also captures interaction metadata --- such as the timing of score selections --- which can be examined later to explore evaluation dynamics and detect potential biases.

\subsubsection{Student Dashboard}
\begin{figure}[t]
  \centering
  \includegraphics[width=0.8\linewidth, height=8cm, keepaspectratio=false]{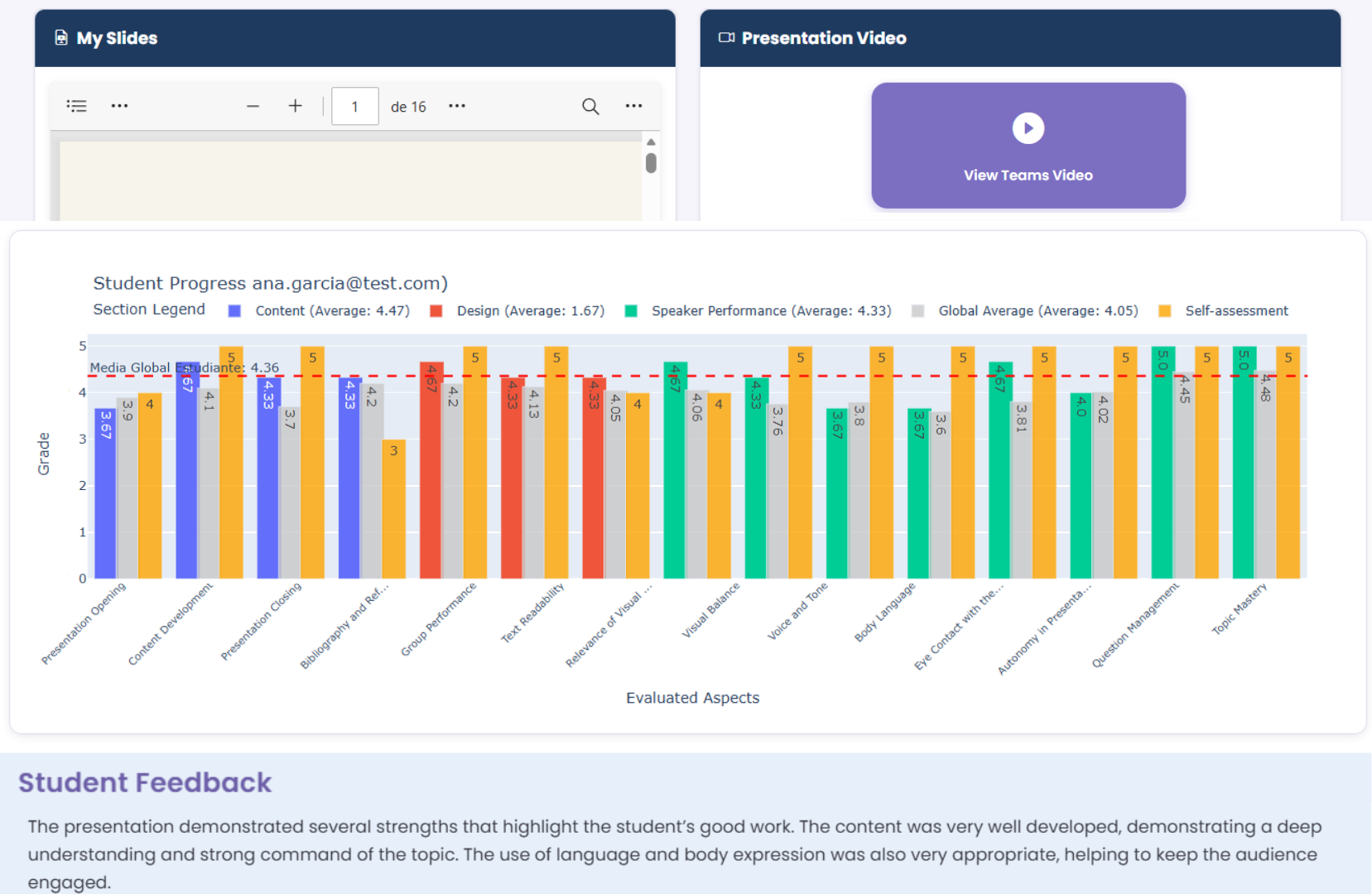}
  \caption{Screenshot of the AICoFe student dashboard.}
  \label{fig:student_dashboard}
\end{figure}

The Student Dashboard (Figure~\ref{fig:student_dashboard}) is designed to foster reflective learning through self-assessment and feedback exploration. Students can access a video recording of their own presentation and complete a self-evaluation using the same rubric applied by peers and instructors. Once submitted, the dashboard visualises differences and alignments between self-assessment scores and the aggregated evaluations received. Students also receive AI-supported collaborative feedback composed by the teacher from the LLM-generated outputs. A short embedded questionnaire allows students to express their agreement with and perceived usefulness of the feedback, providing additional qualitative input for system evaluation.

\subsubsection{Teacher Dashboard}

\begin{figure}[h]
  \centering

  \begin{minipage}{0.8\linewidth}
    \centering
    \includegraphics[width=\linewidth, keepaspectratio]{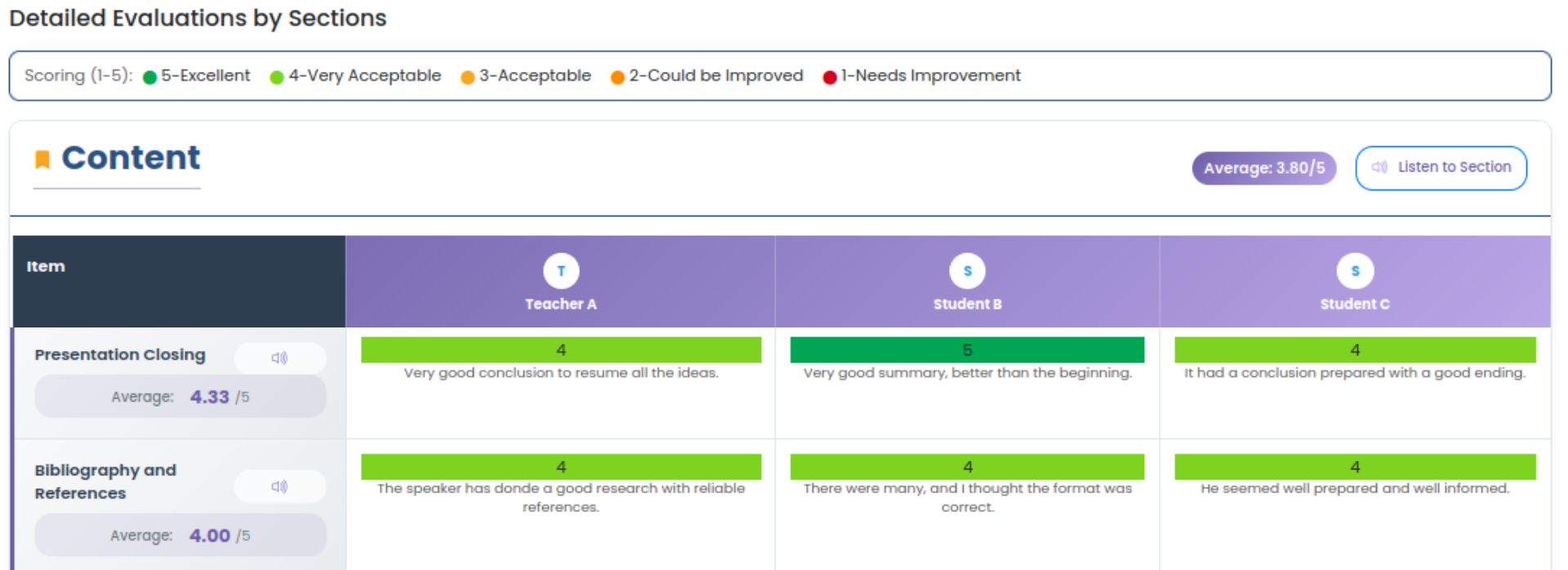}
    \par\small (a)
  \end{minipage}
  
  \vspace{0.6cm}
  
  \begin{minipage}{0.8\linewidth}
    \centering
    \includegraphics[width=\linewidth, keepaspectratio]{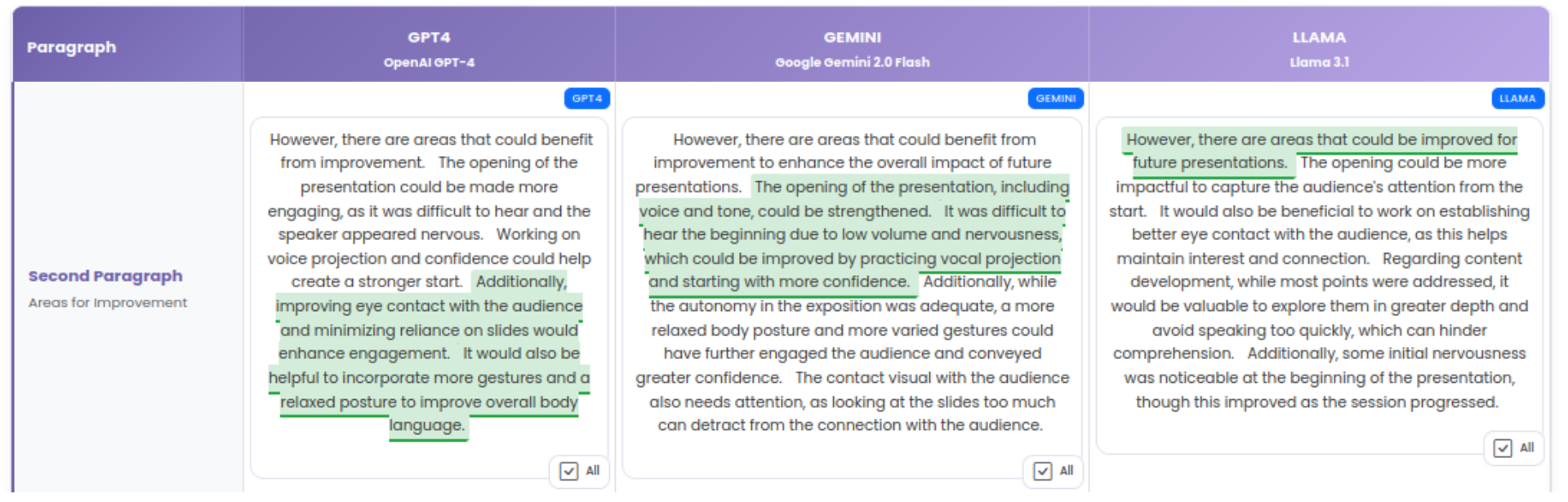}
    \par\small (b)
  \end{minipage}

  \caption{Teacher Dashboard views: (a) granular feedback from the three LLMs with sentence-level selection, and (b) detailed evaluations by rubric item and evaluator.}
  \label{fig:teacher_dash}
\end{figure}

The Teacher Dashboard is the central interface for feedback review and curation. It allows teachers to inspect quantitative scores and qualitative comments provided 
by both peers and teachers (Figure~\ref{fig:teacher_dash}a), access student performance videos, and review the feedback generated by the different LLMs supported by the system. Teachers can also 
compose the final feedback by selecting sentences from the LLM outputs (Figure~\ref{fig:teacher_dash}b).

Teachers can select individual sentences or entire paragraphs from the AI-generated feedback of each LLM and combine them to compose the final feedback that will be presented to the student. This selective composition reinforces the human-centered nature of the system by positioning teachers as active mediators of AI-generated content. Additionally, the dashboard integrates a text-to-speech functionality (using Google Text-to-Speech) that enables teachers to listen to the detailed evaluation comments, facilitating review without the need to read every comment individually.

\subsubsection{Feedback History and LLM Contribution Tracking}

\begin{figure}[h]
  \centering
  \includegraphics[width=\linewidth]{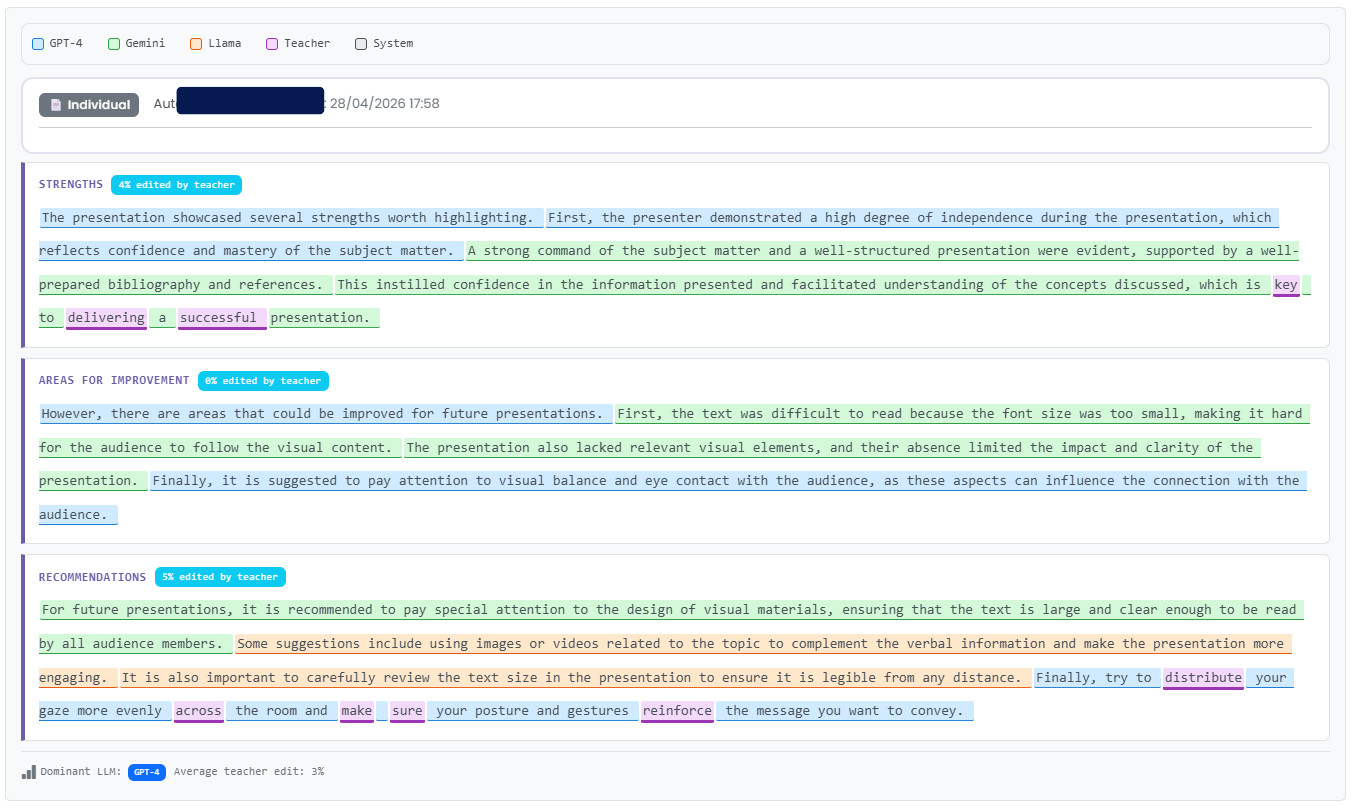}
  \caption{Feedback History interface in the Teacher Dashboard, showing the LLM contribution legend.}
  \label{fig:history}
\end{figure}

A key feature of the Teacher Dashboard is the \textbf{Feedback History} interface (Figure~\ref{fig:history}), which provides a comprehensive log of all feedback sent to each student. For each feedback entry, the interface displays a visual legend indicating the proportion of content contributed by each LLM (GPT-4.1-mini, Gemini 2.5 Flash, and Llama 3.1), as well as the extent to which the teacher modified or curated the AI-generated text before sending it to the student.

This transparency mechanism serves two purposes. First, it allows teachers to reflect on their own curation patterns across students and sessions, potentially identifying tendencies to favour certain LLMs or to apply greater or lesser editorial effort. Second, it provides a data source for analysing the role of teacher mediation in the feedback process at scale.

In addition to the history log, the interface supports draft management: teachers can save intermediate versions of composed feedback as drafts before finalising and sending them to students. This is particularly useful in settings where feedback curation is done across multiple sessions or where the teacher wishes to revisit a draft after reviewing additional student materials.

\subsubsection{Admin Dashboard}

System configuration and evaluation orchestration are managed through the Admin 
Dashboard, which allows administrators to create courses, define student groups, 
assign rubrics, and change the language of the system and the generated feedback. 
This dashboard is intended for system administrators and is not visible to students 
or teachers during normal operation.

\subsection{Feedback Generation Module}

The Feedback Generation Module is responsible for producing personalised feedback by combining quantitative assessment results from teachers and peers with the qualitative observations collected during the evaluation process. The module extends prior work~\cite{becerra2024generative} to a multi-model setting in which three independent feedback instances are generated, one per LLM.

\subsubsection{Pre-processing and Prompt Construction}

As an initial step, the module pre-processes the qualitative input collected from evaluators. Textual comments are normalised to remove invalid characters and are automatically screened to assess their relevance with respect to the evaluated rubric item using a keyword-based filtering heuristic that matches comments against rubric-specific term lists. Once the data have been filtered, the system constructs structured prompts based on course-specific templates. These prompts integrate average scores per rubric item with validated qualitative observations provided by multiple evaluators. To improve contextual accuracy, the prompts also incorporate rubric level descriptions and selected instructional materials associated with the assessed skill. All data sent to the generative models are anonymised prior to transmission, ensuring that no student-identifying information leaves the institutional environment in identifiable 
form before reaching commercial LLM providers.

\subsubsection{Multi-LLM Generation}

The module supports three LLMs: \textbf{OpenAI GPT-4.1-mini}, \textbf{Google Gemini 2.5 Flash}, and \textbf{Meta Llama 3.1}. Using three distinct models enriches pedagogical feedback with diverse perspectives while reducing the risk of single-model biases or systematic errors reaching students.

Each model receives the same structured prompt and independently generates a feedback instance. All three models are fine-tuned using data from the SOPHIAS dataset~\cite{becerra2026sophias}, which comprises 50 AICoFe evaluation instances for which teachers generated feedback using the LLMs and subsequently reviewed and refined it. These teacher-validated feedback instances are used as training data to improve the models' alignment with the pedagogical objectives of the course.

Each feedback instance follows a predefined three-paragraph structure: (i) \textit{strengths}, highlighting well-performed aspects; (ii) \textit{areas for improvement}, identifying specific weaknesses; and (iii) an \textit{action plan}, providing concrete recommendations for future improvement.

Each model was fine-tuned using supervised learning on input--output pairs structured according to the three-paragraph feedback schema. \textbf{GPT-4.1-mini} was adapted via the OpenAI fine-tuning API for 3 epochs
with a learning rate multiplier of 2, consuming 205,239 tokens and reaching a final training loss of 0.934. \textbf{Gemini 2.5 Flash} was fine-tuned through Google Vertex AI for 40 cycles, with the optimal checkpoint automatically
selected at step 45, achieving a final loss of 0.759 and accuracy of 0.798. \textbf{Llama 3.1} was fine-tuned locally via LoRA~\cite{hu2022lora} ($r=16$, $\alpha=32$) with 4-bit NF4 quantisation for 3 epochs, reaching a final training
loss of 1.210. All three models were qualitatively validated through teacher review prior to deployment.

\subsubsection{Post-processing and Validation}

After generation, a validation stage ensures quality. The system verifies constraints defined by teachers, such as minimum and maximum feedback length, overall coherence, and the exclusion of restricted terms (e.g., evaluator names or other identifying information). If the generated feedback does not meet these requirements, the module automatically triggers a regeneration process.

\subsection{Recording Module}

The Recording Module supports reflective learning by capturing students' oral presentations in video and audio format. Recordings are obtained using a standard laptop with an integrated webcam and microphone, relying on Microsoft Teams as the recording platform due to its widespread availability within the institution's infrastructure. Recordings are linked to the corresponding evaluation instances in the database and are made accessible to both students (for self-assessment) and teachers (for evaluation review).

\section{Conclusions and Future Work}
\label{sec:conclusions}

This paper has described the implementation and deployment of AICoFe, an AI-based collaborative feedback system designed to support peer, self-, and teacher assessment in higher education. We have detailed the system's modular architecture, including its multi-LLM feedback generation pipeline, role-specific Learning Analytics dashboards, and feedback history interface with LLM contribution tracking and draft management. Evaluation results confirm that students perceive the AI-assisted feedback as highly coherent and useful, while system usability scores indicate excellent acceptance across both undergraduate and master's levels.

AICoFe is currently being deployed at a larger scale during the ongoing academic year, involving approximately 80 students and 5 teachers across one undergraduate course and one master’s course at Universidad Autónoma de Madrid (UAM). This extended deployment will allow a more comprehensive evaluation of the system’s impact on peer feedback quality and reflective learning. In particular, future analyses will examine students' perceptions of feedback utility, coherence, and actionability, compare AI-mediated feedback with traditional manual feedback practices where possible, and investigate whether the use of AICoFe is associated with improvements in feedback quality, reflective learning, and presentation performance. Future developments will also explore the integration of multimodal data, including non-verbal cues captured during presentations into the feedback generation process.~\cite{becerra2025mosaic,becerra2025enhancing_biosensors,golrang2025does}.

\section*{Acknowledgements}
    Support by projects: Cátedra ENIA UAM-VERIDAS en IA Responsable (NextGenerationEU PRTR TSI-100927-2023-2), M2RAI (PID2024-160053OB-I00, MICIU/FEDER) and SNOLA (RED2022-134284-T). Alvaro Becerra is funded by a predoctoral contract (FPI) from the Comunidad de Madrid (PIPF-2024/COM-34288).

\bibliography{lasi2026}


\end{document}